\documentclass[12pt,a4paper]{article}

\usepackage{epsfig,cite,amsmath,authblk}

\title{Considering Air Density in Wind Power Production}

\author{Z\'en\'o Farkas}

\affil{Dept. of Physics of Complex Systems, E\"otv\"os University, P\'azm\'any
  P. stny 1A, H-1117 Budapest, Hungary}

\date{2011}

\begin{document}

\maketitle

\begin{abstract}
  In the wind power production calculations the air density is usually
  considered as constant in time.  Using the CIPM-2007 equation for the
  density of moist air as a function of air temperature, air pressure and
  relative humidity, we show that it is worth taking the variation of the air
  density into account, because higher accuracy can be obtained in the
  calculation of the power production for little effort.
\end{abstract}

Wind forecasts for wind energy applications rely mostly on wind speed and
direction, and only marginally on the forecast of air density
\cite{monteiro2009}. In the wind power literature, the air density is usually
considered as constant in time, with standard value $\rho =
1.225\:\mathrm{kg/m^3}$ (at sea level, $15\:^\circ\mathrm{C}$)
\cite{weisser2003wind,jamil1995wind,manwell2010wind,gokcek2007investigation},
in the formula for the wind power power production:
\begin{equation}
  P = \frac{1}{2} \rho A v^3,
  \label{eq:powerproduction}
\end{equation}
where $A$ is the area swept by the turbine blades and $v$ is the wind
speed. In this paper we argue that it is worth considering the variation of
the air density in the calculation of the power production, because higher
accuracy can be obtained.

For the calculation of air density as a function of air temperature $t$
[$^\circ$C], air pressure $p$ [Pa] and relative humidity $h$ $(0 \leq h \leq
1)$, we use the current official formula of the International Committee for
Weights and Measures (CIPM), referred to as CIPM-2007 equation
\cite{CIPM2007}:
\begin{equation}
  \rho(t,p,h) = \frac{p M_\mathrm{a}}{Z(t,p,h) R T(t)}
  \left\{1 - x_\mathrm{v}(t,p,h)
    \left[1 - \frac{M_\mathrm{v}}{M_\mathrm{a}}\right]\right\},
  \label{eq:CIPM2007}
\end{equation}
where $R = 8.314\,472\:\mathrm{J} / \mathrm{mol\:K}$ is the molar gas
constant, $T(t) = (273.5 + t/^\circ\mathrm{C})\;\mathrm{K}$ is the
thermodynamic temperature, $M_\mathrm{a} = 28.965\,46 \times 10^{-3}\;
\mathrm{kg}/\mathrm{mol}$ is the molar mass of dry air, $M_\mathrm{v} =
18.015\,25 \times 10^{-3}\;\mathrm{kg}/\mathrm{mol}$ is the molar mass of
water, $x_\mathrm{v}(t,p,h)$ is the mole fraction of water vapour:
\begin{equation}
  x_\mathrm{v}(t,p,h) = h\;[\alpha + \beta p + \gamma t^2]\;
  \frac{\exp\left[A T(t)^2+B T(t)+C+\frac{D}{T(t)}
      \right]\mathrm{Pa}}{p},
  \label{eq:vapourfrac}
\end{equation}
and $Z(t,p,h)$ is the compressibility factor:
\begin{multline}
  Z(t,p,h) = 1 - \frac{p}{T(t)}\;
  \left\{a_0+a_1t+a_2t^2+[b_0+b_1t]
  x_\mathrm{v}+[c_0+c_1t]x_\mathrm{v}^2\right\} \\
  + \frac{p^2}{T(t)^2}\;[d+e x_\mathrm{v}^2].
  \label{eq:compressibility}
\end{multline}
The constants in Eqs. (\ref{eq:vapourfrac}) and
(\ref{eq:compressibility}) are:

\begin{eqnarray*}
  A &=& 1.237\,884\,7 \times 10^{-5} \:\mathrm{K}^{-2} \\
  B &=& −1.912\,131\,6 \times 10^{-2} \:\mathrm{K}^{-1} \\
  C &=& 33.937\,110\,47 \\
  D &=& -6.343\,164\,5 \times 10^3 \:\mathrm{K} \\
  \alpha &=& 1.000\,62 \\
  \beta &=& 3.14 \times 10^{-8} \:\mathrm{Pa}^{-1} \\
  \gamma &=& 5.6 \times 10^{-7} \:\mathrm{K}^{-2} \\
  a_0 &=& 1.581\,23 \times 10^{-6} \:\mathrm{K} \:\mathrm{Pa}^{-1} \\
  a_1 &=& -2.9331 \times 10^{-8} \:\mathrm{Pa}^{-1} \\
  a_2 &=& 1.1043 \times 10^{-10} \:\mathrm{K} ^{-1} \:\mathrm{Pa}^{-1} \\
  b_0 &=& 5.707 \times 10^{-6} \:\mathrm{K} \:\mathrm{Pa}^{-1} \\
  b_1 &=& -2.051 \times 10^{-8} \:\mathrm{Pa}^{-1} \\
  c_0 &=& 1.9898 \times 10^{-4} \:\mathrm{K} \:\mathrm{Pa}^{-1} \\
  c_1 &=& -2.376 \times 10^{-6} \:\mathrm{Pa}^{-1} \\
  d &=& 1.83 \times 10^{-11} \:\mathrm{K} 2 \:\mathrm{Pa}^{-2} \\
  e &=& -0.765 \times 10^{-8} \:\mathrm{K} 2 \:\mathrm{Pa}^{-2}
  \label{eq:constants}
\end{eqnarray*}

\begin{figure}
  \begin{center}
    \includegraphics[angle=0,width=12cm]{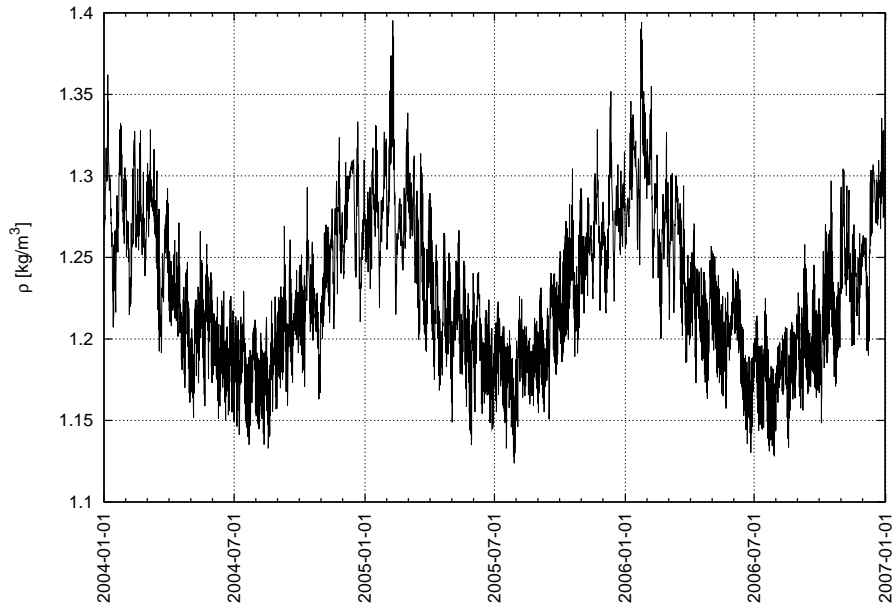}
    \caption{Air density in Gy\H{o}r, Hungary from 2004 to 2006, as calculated
      from the CIPM-2007 equation (\ref{eq:CIPM2007}). The temperature,
      pressure and relative humidity data were measured at the local station
      of the Hungarian Meteorological Service.}
    \label{fig:AirDensityGyor}
  \end{center}
\end{figure}

The wind speed and power data presented here were measured between 2004 and
2006 at an Enercon E-40 wind turbine of rated power 600 kW, near Mosonszolnok,
Hungary, installed and operated by E.ON Hungary. The meteorological data
necessary to calculate the air density were obtained from a nearby
meteorological station, Gy\H{o}r, for the same time interval. The air density
calculated according to the CIPM-2007 equation (\ref{eq:CIPM2007}) is shown in
Fig.~\ref{fig:AirDensityGyor}. The average density was
$\rho = 1.229\:\mathrm{kg/m^3}$, which is rather close to the standard
$\rho = 1.225\:\mathrm{kg/m^3}$ value.

\begin{figure}
  \begin{center}
    \includegraphics[angle=0,width=10cm]{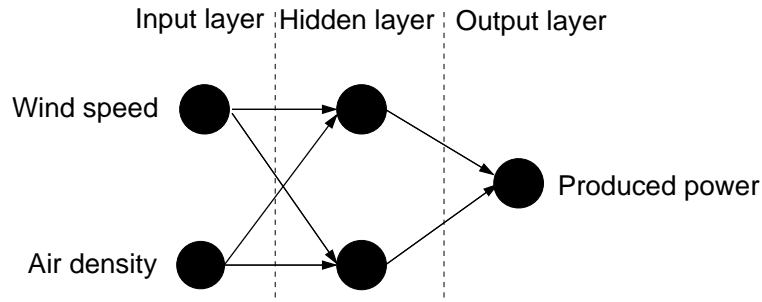}
    \caption{The topology of the neural network used for power curve fitting.}
    \label{fig:NeuralNetwork}
  \end{center}
\end{figure}

For fitting the power curve for the Enercon E-40 turbine we used a very simple
neural network \cite{dreyfus2005neural} with two input nodes (one for the wind
speed, the other for the air density), two nodes in the hidden layer, and one
output node for the power. The transfer function was $\tanh(\cdot)$, which,
due to the general shape of the power curve, seemed especially well suited for
this task. Fig.~\ref{fig:NeuralNetwork} shows the configuration of this simple
neural network.

\begin{figure}
  \begin{center}
    \includegraphics[angle=0,width=12cm]{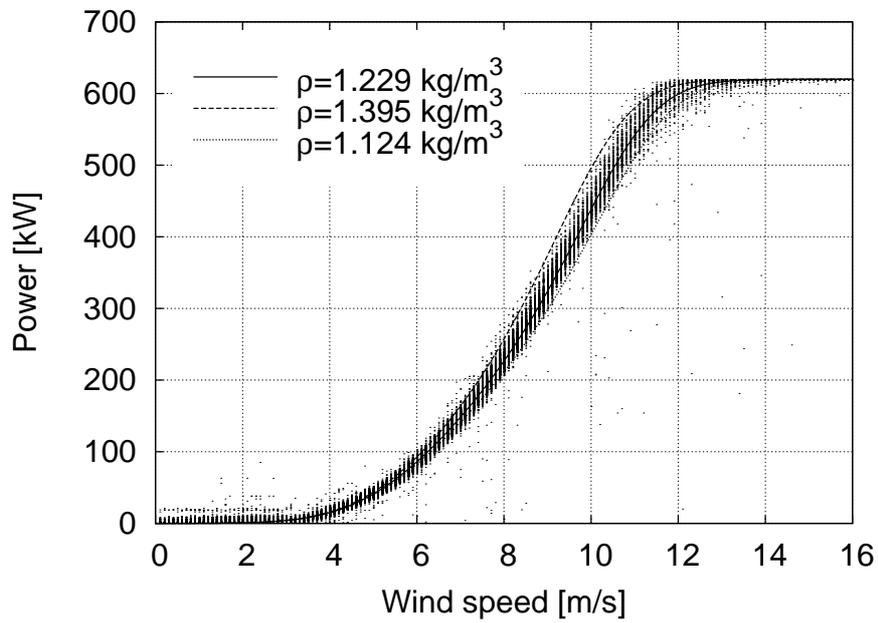}
    \caption{The power produced by the Enercon E-40 turbine in 2006 as a
      function of the wind speed (dots) and the curves fitted to the
      2004--2005 data using neural network model for the average air density
      $\rho=1.229\:\mathrm{kg}/\mathrm{m}^3$ (dashed line), at high density
      $\rho=1.395\:\mathrm{kg}/\mathrm{m}^3$ (dashed line), and at low density
      $\rho=1.124\:\mathrm{kg}/\mathrm{m}^3$ (dotted line).}
    \label{fig:PowerSpeed}
  \end{center}
\end{figure}

Fig.~\ref{fig:PowerSpeed} shows the measured and fitted power curve for this
wind turbine. For the fitting we used data from the first two years, 2004 and
2005, and evaluated the results for the data from 2006 (out-of-sample
fitting). When the air density is kept constant at the average value
$\rho=1.229\:\mathrm{kg}/\mathrm{m}^3$, the RMS error of the fitted power curve
is $12.06$~kW. In contrast, with the full fitting model (having the air
density as the second input in the neural network) the RMS error is
$10.15$~kW, which $16\%$ lower.

In conclusion, we found that taking air density into account in determining
wind power production decreases the potential RMS error by up to $16\%$. This
also means that if air density is not considered, a systematic error of this
magnitude is introduced in the wind power calculation. Therefore we strongly
suggest considering air density in wind power forecasting.

This work has been supported by the European Union and co-financed by the
European Social Fund (grant agreement no. TAMOP 4.2.1./B-09/1/KMR-2010-0003).

\bibliography{air_density_paper}{}
\bibliographystyle{ieeetr}

\end{document}